 \documentclass[
twocolumn,showpacs, amsmath, amssymb, floatfix,
nofootinbib,wrapfloat byrevtex]{revtex4}

 \usepackage{amsmath,mathrsfs,graphicx,epstopdf,placeins,float}
 \usepackage{booktabs}
 \usepackage{multirow}
 \usepackage{pstricks}

 \newcommand{\beq}[1]{\begin{equation}\label{#1}}
 \newcommand{\eeq}{\end{equation}}
 \newcommand{\bea}[1]{\begin{eqnarray}\label{#1}}
 \newcommand{\eea}{\end{eqnarray}}

\begin{document} 

 \title{Brane universe and holography in spacetime of charged AdS dilaton black hole}
 \author{Ai-chen Li $^{a,b}$ }
 \email{lac@emails.bjut.edu.cn}
 \email{aichenli@visit.uaveiro.eu}
 \affiliation{\it ${}^a$ Theoretical Physics Division, College of Applied Sciences, Beijing University of Technology }
 \affiliation{\it ${}^b$ Departamento de matematica da Universidade de Aveiro and CIDMA,\\ Campus de Santiago, 3810-183 Aveiro, Portugal} 
 
 \begin{abstract}
In the background of a charged AdS dilaton black hole, we investigate the movement of a self-graviting 3-brane and relevant holographic effects as the brane move close to the AdS boundary. The induced metric on brane corresponds to an exact FLRW geometry, while the evolution of brane is determined by Israel junction condition and the effective Einstein field equation on brane together. When the brane approaches the AdS boundary, AdS/CFT correspondence implies that a radiation dominated FLRW-universe ($P=\frac{1}{3}\rho$) should be given. According to the holographic renormalization procedure, we involve an appropriate surface counterterm into the gravitational action for achieving  $P=\frac{1}{3}\rho$ on brane. This surface counterterm also plays a important role in caculating the mass of charged AdS dilaton black hole. Finally, we obtain the thermodynamic quantities and give an extend Cardy-Verlinde formula on brane. 
 \end{abstract}
\maketitle

\section{Introduction}
The study about physics beyond 4-dimensional spacetime has made great progress in recent decades, a lot of interesting physics was obtained from either the phenomenological aspects or theoretical implications, such as M theory \cite{Horava:1995qa,Horava:1996ma}, Anti de-Sitter space/conformal field theory correspondence \cite{Maldacena:1997re,Witten:1998qj} (i.e. AdS/CFT), and brane world scenario in which our universe is viewed as a 3+1 dimensional brane embedded in a higher dimensional spacetime \cite{Randall:1998uk,Randall:1999ee,Randall:1999vf}. Actually, the original concept of brane could be traced back to the work of Rubakov and Shaposhnikov \cite{Rubakov:1983bb,Rubakov:1983bz}, it was also developed to show that the fundamental gravitional scale could be close to Tev sacle by introducing large extra spatial dimension transverse to the brane \cite{ArkaniHamed:1998rs,Antoniadis:1998ig,ArkaniHamed:1998nn}. Compared to the previous ones, an important variant of barne world scenario involves warped compactification, this idea was first considered by RS-I model \cite{Randall:1999ee} in order to solve the gauge hierarchy problem, which introduced a 5-dimensional AdS compactified extra dimension between two Minkowski branes with a exponential warp factor in metric solution. For implementing localization of 4D graviton and reproducing Newton's law on the visible brane, RS-II model was proposed, which has a noncompactified extra dimension. From perspective of phenomenology, unlike the ADD model, these warped models have less striking signature at colliders or in astrophysical processes, and concomitantly being less constrained. In cosmological applications of brane world, RS-II model has been most successful model since it provided the capability of modifying the structure of Einstein's field equations \cite{Shiromizu:1999wj,Bostock:2003cv}.

For applying brane world to cosmology, a time dependent solution need to be constructed. There exist two physical scenarios, one focus on finding time dependent bulk geometry as cosmological solution \cite{Kaloper:1999sm,Kanti:1999sz,Flanagan:1999cu,Kanti:2018ozd}, another one consider a alternative case in which the bulk remains to be static but the brane acquires a velocity, while the observers on the brane will feel cosmological expansion or contraction as the brane moves \cite{Chamblin:1999ya,Kraus:1999it,Creek:2006je}. Actually, some enriched physical phenomena will be uncovered when considering the movement of brane in AdS spacetime. More interesting, if one considers the AdS black hole as bulk spacetime for a motional brane, the physics of black hole, brane universe and holography could be connected together naturally in this way. Under the situation of a motional brane in spacetime of AdS-Schwarzchild black hole, associating with the AdS/CFT correspondence and black hole thermodynamics, \cite{Savonije:2001nd} generalize the Cardy formula \cite{Cardy:1986ie} to the arbitrary-dimensional spacetime. Inspired by this work, the generalized Cardy formula (called by Cardy-Verlinde formula) have been checked in various black holes bulk with AdS asymptotics \cite{Klemm:2001db, Cai:2001jc, Birmingham:2001vd, Youm:2001yq, Cai:2001ja, Youm:2001qr, Jing:2002aq, Cai:2002bn, Lee:2008yqa, Nojiri:2000pi, Nojiri:2001fa, Nojiri:2002hz, Lidsey:2002ah, Nojiri:2002td, Nojiri:2002vu, BravoGaete:2017dso}. Besides, \cite{Kawai:2015lja} constructs a holographic reheating model by considering the movement of a probe brane in AdS-Vaidya spacetime, and the formation process of AdS black hole dual to the thermalization process of brane universe. Recently, through constructing a model of moving brane in AdS-Schwarzchild black hole, the holographic complexity growth in a FLRW universe is considered by \cite{An:2019opz}. 

As the low-energy effective theory of supergravity and string \cite{Gibbons:1987ps}, dilaton gravitation theories have attracted many attentions in recent decades. Inspired by AdS/CFT, many works devote to finding the asymptotically AdS black hole solution in dilaton gravity with various self-interacting potential of dilaton. Particularly, it is of great interest to consider Liouville-type dilaton potential which is originated from the supersymmetry breaking of a higher-dimensional supergravity model \cite{Gregory:1992kr,Horne:1992bi,Poletti:1994ff}. And \cite{Poletti:1994ff,Mignemi:1991wa} has proved that in models of one and two Liouville-type potential, there doesn't exist the asymptotically flat or asymptotically AdS black hole solution. By combining three different Liouville-type dilaton potentials, \cite{Gao:2004tu,Gao:2004tv,Gao:2005xv} obtains the asymptotically AdS black hole solution in Einstein-Maxwell-Dilaton theory. Basing on this black hole solution, many interesting physical phenomenology have been explored, like the black hole thermodynamics \cite{Sheykhi:2009pf,Sheykhi:2016syb}, holographic thermalization \cite{Zhang:2015dia}, black hole phase transition in extend phase space \cite{ Li:2017kkj}, and domain wall cosmology \cite{Xu:2019abl}.

In this paper, our purpose is to consider the movement of brane in the charged AdS dilaton black hole solved by \cite{Gao:2004tv,Gao:2005xv} and the relevant holographic effects when brane approaches the AdS boundary. As showed in \cite{Sheykhi:2009pf,Li:2017kkj}, a rich phases structure of black hole thermodynamics could be uncovered when adjusting the dilaton coupling constant. Meanwhile, when considering the motion of domain wall in this charged AdS dilaton black hole, \cite{Xu:2019abl} also observes enriched evolution modes of domain wall universe as varying the value of dilaton coupling constant. Hence, it seems that there exists a connection between the thermodynamics of AdS dilaton black hole in bulk spacetime and the evolution of FLRW universe on brane/wall. Basing on the idea proposed in \cite{Savonije:2001nd}, namely associating the black hole thermodynamics with AdS/CFT correspondence, \cite{Cai:2001jc} obtains a modified Cardy-Verlinde formula for a CFT living in the boundary of AdS-RN black hole spacetime. In term of the AdS dilaton black hole \cite{Gao:2004tv,Gao:2005xv}, when we make the dilaton coupling constant $\alpha=0$, this solution will reduce to the AdS-RN one accordingly. Thus, when brane approaches the AdS boundary, we naturally expect to give a similiar Cardy-Verlinde formula on brane like the one obtained by \cite{Cai:2001jc} but with some corrections of dilaton coupling constant $\alpha$. Note that the matter field confined on brane is a general quantum field theory (QFT) without conformal symmetry \cite{Boonstra:1998mp,Behrndt:1999mk,Cvetic:2000pn}. And we need to involve an appropriate surface counterterm to restore the conformal symmetry when brane approaches the boundary of AdS dilaton black hole. Besides, although \cite{Xu:2019abl} have considered the movement of brane/wall in AdS dilaton black hole \cite{Gao:2004tv,Gao:2005xv}, they ignore the self-gravitating effects of brane/wall. Thus, in their case, the evolution of brane/wall is controlled by Israel junction condition \cite{Israel:1966rt} only. However, in our scenario, we involve the gravity on brane by using the method provided in \cite{Shiromizu:1999wj}, and the evolution of brane is determined by the effective Einstein field equation on brane and Israel junction condition together.

Our work is organized as follows. In Sec.\ref{BHpart}, we briefly review the 5-dimensional asymptotically AdS black hole solution and relevant thermodynamics. Besides, we give a well-defined boundary stress-energy tensor and caculate the black hole mass  by adding an appropriate surfacte counterterms to the gravitational action. The movement of self-graviting brane in the background of the AdS dilaton black hole are considered in Sec.\ref{BranMotion}. In Sec.\ref{BranHolo}, the holographic effects on brane will be investigated as the brane approaches the AdS boundary. Finally, Sec.\ref{ConAndDis} will summarize our results and give a discussion.

\section{AdS dilaton black holes and relevant thermodynamical quantities \label{BHpart}}

\subsection{Dilaton black holes solution in asymptotically AdS spacetime}
\begin{figure}[ht]
	\begin{center}
		\includegraphics[scale=0.75]{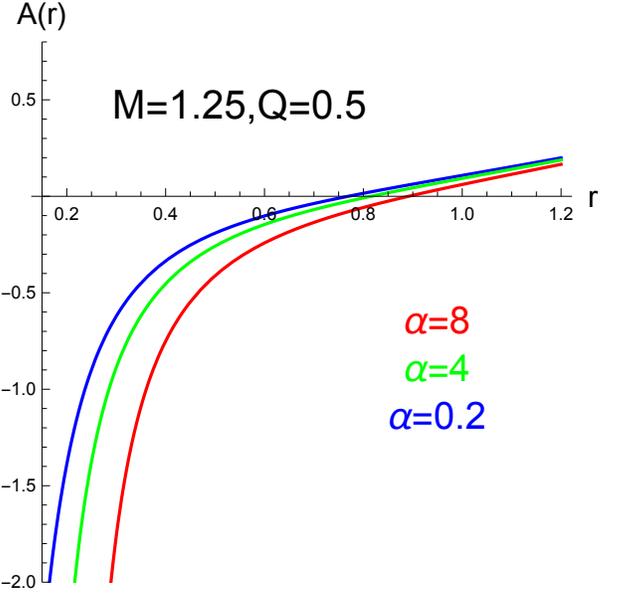}
		\caption{(color online). Plot the horizon function $A(r)$ of charged AdS dilaton black hole solution at fixed black hole mass and charge with different dilaton coupling constant..}
		\label{figAdShorizon}
	\end{center}
\end{figure}

A asymptotically AdS dilatonic black hole with charge will be considered as a background spacetime (i.e. bulk) for brane's motion, thus we will review this black hole solution in present section. We begin with the action of 5-dimensional Einstein-Maxwell-dilaton gravity (we use the convention $\kappa^2=8\pi G$)
\bea{EMD}
&&\hspace{-5mm}S_{EMD}=\frac{1}{2\kappa_{5}^{2}}\int_{M}d^{5}x\sqrt{-g}[\mathcal{R}-\frac{4}{3}g^{MN}\partial_{M}\phi\partial_{N}\phi
\\
&&\rule{15mm}{0pt}-V(\phi)-e^{-\frac{4}{3}\alpha\phi}F^{2}]
\nonumber
\eea
where $\mathcal{R}$ and $\phi$ are the usual Ricci scalar and dilaton field respectively, the latter has self-interaction $V(\phi)$ and non-minimally couples to the electromagnetic field of kinetic energies $F^2$. The physical constants $\alpha$ measures the strength of this coupling. Equation of motions following from this action have the form
\bea{feq}
\nonumber
&&\hspace{-5mm}\mathcal{R}_{MN}=\frac{1}{3}[4\partial_{M}\phi\partial_{N}\phi+g_{MN}V(\phi)]+2e^{-\frac{4\alpha\phi}{3}}[F^L _M F_{LN}                               
\\
&&\hspace{-5mm} \quad \quad \quad -\frac{1}{6}g_{MN}F^{2}] 
\eea
\bea{seq}
&&\hspace{-11mm} \nabla^{2}\phi=\frac{\partial_{M}(\sqrt{-g}g^{MN}\partial_{N}\phi)}{\sqrt{-g}}=\frac{3}{8}\frac{\partial V}{\partial\phi}-\frac{\alpha}{2}e^{-\frac{4\alpha\phi}{3}}F^{2}  
\eea
\bea{Maxwell}
&&\hspace{-11mm} \nabla_{N}(e^{-\frac{4\alpha\phi}{3}}F^{NM})=\partial_{N}(\sqrt{-g}e^{-\frac{4\alpha\phi}{3}}F^{NM})=0
\eea
We consider a static black hole solution with metric ansatz
\bea{Metric}
\nonumber
&& ds^2 =g_{AB}dx^Adx^B \\
\label{Ge5dBHAnsa}
&& \quad ~=-A(r)dt^2+B(r)dr^2+R(r)^2d\Omega ^2 _{k,3}
\eea
where $d\Omega ^2 _{k,3}$ is the line element of 3-dimensional hyper surface of constant curvature $6k$ with $k=\pm1, 0$ corresponding to spheric, hyperbolic and plane topology respectively. In the case of only static electric fields occcur in the system, the only nonzero components of $F_{M N}$ could be obtained from the maxwell equation \eqref{Maxwell}
\bea{solEM}
F_{tr} =\sqrt{A(r)B(r)}\frac{qe^{4\alpha\phi/3}}{R^{3}(r)}
\eea
Explicitly, substitute \eqref{Metric},\eqref{Maxwell} into \eqref{feq} and \eqref{seq}, it can be obtained that
\bea{}
\label{Eintt}
&&\hspace{-14mm}\frac{A''}{2B}+\frac{3A'R'}{2BR}-\frac{A'B'}{4B^{2}}-\frac{(A')^{2}}{4AB}=\frac{4e^{-4\alpha\phi/3}(F_{tr})^{2}}{3B}-\frac{1}{3}AV
\\
\label{Einrr}
&&\hspace{-14mm}\frac{A''}{2A}+\frac{3R''}{R}-\frac{3B'R'}{2BR}-\frac{A'B'}{4AB}-\frac{(A')^{2}}{4A^{2}}=-\frac{1}{3}BV-\frac{4}{3}(\phi')^{2}\\
\nonumber
&&\hspace{-4mm}\quad\quad\quad\quad\quad\quad\quad\quad\quad\quad\quad\quad\quad\quad +\frac{4(F_{tr})^{2}e^{-4\alpha\phi/3}}{3A}
\\
\label{Einxx}
&&\hspace{-14mm}\frac{RR''}{B}+\frac{2(R')^{2}}{B}-\frac{B'R'R}{2B^{2}}+\frac{A'R'R}{2AB}=-\frac{1}{3}R^{2}V
\\
\nonumber 
&&\hspace{-4mm}\quad\quad\quad\quad\quad\quad\quad\quad\quad\quad\quad\quad\quad-\frac{2R^{2}(F_{tr})^{2}e^{-4\alpha\phi/3}}{3AB}
\\
\label{eqphi}
&&\hspace{-14mm}\frac{\phi''}{B}+\frac{3R'\phi'}{BR}-\frac{B'\phi'}{2B^{2}}+\frac{A'\phi'}{2AB}=\frac{3}{8}V'+\frac{\alpha(F_{tr})^{2}e^{-4\alpha\phi/3}}{AB}
\eea
where we have set $k=1$ with topology $S^3$, namely $d\Omega ^2 _3 =d\theta ^2 _1+\sin^2\theta_1 d\theta^2 _2+\sin^2\theta_1 \sin^2 \theta_2 d\varphi ^2$. For simplicity, we will consider only the spherical case in this paper. By adjusting the form of $V(\phi)$ appropriately, reference \cite{Gao:2004tv, Gao:2005xv} obtains asymptotically AdS black hole solutions of the system analytically,
\bea{}
\label{solDilaPon}
&&\hspace{-3mm}V(\phi)=\frac{\Lambda}{2(2+\alpha^{2})^{2}}(4\alpha^{2}(\alpha^{2}-1)\cdot e^{-\frac{8\phi}{3\alpha}}\\
\nonumber
&& \hspace{-3mm}\quad \quad ~ +4(4-\alpha^{2})\cdot e^{\frac{4\alpha\phi}{3}}+24\alpha^{2}\cdot e^{-\frac{2(2-\alpha^{2})\phi}{3\alpha}})
\\
\label{solA}
&&\hspace{-3mm}A(r)=-\frac{c^{2}}{r^{2}}(1-\frac{b^{2}}{r^{2}})^{1-\frac{2\alpha^{2}}{2+\alpha^{2}}}-\frac{\Lambda r^{2}}{6}(1-\frac{b^{2}}{r^{2}})^{\frac{\alpha^{2}}{2+\alpha^{2}}}
\\
\label{solB}
&&\hspace{-3mm}B(r)=(1-\frac{b^{2}}{r^{2}})^{-\frac{\alpha^{2}}{2+\alpha^{2}}}/A(r)
\\
\label{solR}
&&\hspace{-3mm}R(r)=(1-\frac{b^{2}}{r^{2}})^{\frac{\alpha^{2}}{2(2+\alpha^{2})}}r
\\
\label{solPhi}
&& \hspace{-3mm}\phi(r)=\frac{3\alpha}{2(2+\alpha^{2})}\ln(1-\frac{b^{2}}{r^{2}})
\eea
Besides, the following relation is also implied by Einstein field equations,
\bea{}
\label{ElectriWithbc}
q^{2}=\frac{6}{(2+\alpha^{2})}b^{2}c^{2}
\eea
Here $b$ and $c$ are integration constants with dimension of length, as we show later, they are also related with the mass $M$. Reference \cite{Sheykhi:2009pf} shows that for this black hole solution, the Kretschmann scalar $R^{\mu \nu \alpha \beta} R_{\mu \nu \alpha \beta}$ and the Ricci scalar R both diverge at $r =b$, thus $r =b$ is the location of curvature singularity. Furthermore, as shown in Fig.\ref{figAdShorizon}, for AdS black hole solution, there only exists one horizon whatever the value of dilaton coupling constant $\alpha$.

\subsection{the mass of black holes}

When we consider the black hole thermodynamics through Euclidean path integral approach, there exists a unavoidable divergence on boundary of spacetime. In asymptotically AdS spacetime, a suitable surface counterterm could be found with feature of coordinate frame independence, after the renormalization procedure, a finite Euclidean action and a well-defined boundary stress-energy tensor will be obtained. According to the methods of Ref.\cite{Balasubramanian:1999re,Dayyani:2016gaa,Cai:1999xg}, we choose the surface counterterm as the following ansatz
\bea{}
\nonumber
&&\hspace{-2mm} S_{ct}	=	-\frac{1}{\kappa_{5}^{2}}\int_{\partial M}d^{4}x\sqrt{-\gamma}\bigg\{\frac{c_{0}}{l_{eff}(\phi)}(1+\frac{c_{\phi}}{c_{0}}\phi^{2})\\
\label{CounterAction}
&&\hspace{-2mm}\quad \quad \quad +c_{1}l_{eff}\mathcal{R}+c_2l_{eff}^{3}\big(\mathcal{R}^{2}+c_\beta \mathcal{R}^{ab}\mathcal{R}_{ab}\big)\bigg\} \\
\nonumber
\\
\nonumber
&& where\quad  \frac{1}{l_{eff}(\phi)}=\sqrt{-\frac{V(\phi)}{3(3+1)}} 
\eea
the expression of $V(\phi)$ has shown in eqs.$\eqref{solDilaPon}$. Actually, $\mathcal{R}^{2}$ and $\mathcal{R}^{ab}\mathcal{R}_{ab}$ are similiar in magnitude, we will make $c_\beta=0$ for simplification in caculation. With inclusion of this counterterm, the quasilocal stress-energy tensor at the boundary $r=const$ with induced metric $\gamma_{ab}$ could be derived as
\bea{}
\nonumber
&&T_{ab}	=	\frac{1}{\kappa_{5}^{2}}\bigg\{ K_{ab}-K\gamma_{ab}+\frac{c_{0}}{l_{eff}}\big(1+c_{\phi}\phi^{2}\big)\gamma_{ab}\\
\nonumber
&&\quad \quad -2c_{1}l_{eff}\big(\mathcal{R}_{ab}-\frac{1}{2}\mathcal{R}\gamma_{ab}\big)+c_{2}l_{eff}^{3}\big(\gamma_{ab}\mathcal{R}^{2}\\
\label{GeneTab}
&&\quad \quad-4\mathcal{R}\mathcal{R}_{ab}+4\nabla_{a}\nabla_{b}\mathcal{R}-4\gamma_{ab}\nabla_{m}\nabla^{m}\mathcal{R}\big)\bigg\}		
\eea
where the $\gamma_{ab}$ is the induced metric on the boundary $r=const$, which is defined as
\bea{}
\gamma_{ab}dx^adx^b=\lim_{r\to con} ds^2_5=-A(r)dt^2 +\big(R(r) \big) ^2d\Omega^2_3
\eea
The $K_{ab}$ is the extrinsic curvature on the boundary. Expand the $\eqref{GeneTab}$ explicitly, we give
\bea{}
\nonumber
&&T_{tt}=-3\frac{AR^{\prime}}{\sqrt{B}R}-\frac{c_{0}}{l_{eff}}A\big(1+c_{\phi}\phi^{2}\big)\\
\label{WithCouTtt}
&&\quad \quad -6\frac{c_{1}l_{eff}}{R^{2}}A-36\frac{c_{2}l_{eff}^{3}}{R^{4}}A\\
\nonumber
&&T_{ij}	=	\frac{h_{ij}}{R^{2}}\bigg(\frac{R^{2}A^{\prime}}{2A\sqrt{B}}+2\frac{RR^{\prime}}{\sqrt{B}}+c_{0}(1+\frac{c_{\alpha}}{c_{0}}\phi^{2})\frac{R^{2}}{l_{eff}}\\
\label{WithCouTij}
&& \quad \quad +2c_{1}l_{eff}-12c_{2}\frac{l_{eff}^{3}}{R^{2}}\bigg)
\eea
The mass of black hole is a conserved charge associated with a timelike killing vector, from the reference \cite{Balasubramanian:1999re}, it could be defined as
\bea{}
\label{DefiOfBHM}
M=\int _{r\to \infty} dx^3  \big(R(r) \big) ^3 \big( A(r) \big) ^{-\frac{1}{2}} T_{tt}
\eea 
For getting a finite mass of black hole, we need to find a well-defined quasilocal stress-energy $T_{tt}$. Thus, we choose the undetermined coefficients in \eqref{CounterAction} as $c_0=-3, c_\phi=\frac{2}{9}, c_1=-\frac{1}{4}, c_2=\frac{1}{96}$. After substitute these coefficients and \eqref{solDilaPon}-\eqref{solPhi} into $\eqref{DefiOfBHM}$, we obtain
\bea{}
\label{FiExOf}
M=\frac{3\Omega_{3}}{2\kappa_{5}^{2}}\bigg(c^{2}+\big(\frac{2-\alpha^{2}}{2+\alpha^{2}}\big)b^{2}\bigg)
\eea

\subsection{the thermodynamical quantities of black hole}

In term of the general metric ansatz $\eqref{Ge5dBHAnsa}$, the Hawking temperature could be expressed as
\bea{}
\label{DefiTema}
T_{H}=\frac{\sqrt{A^{\prime}(B^{-1})^{\prime}}}{4\pi}\vert_{r=r_{+}}
\eea
Substitute \eqref{solA} and \eqref{solB} into \eqref{DefiTema}, we obtain
\bea{}
\nonumber
&&T_H=\frac{\big((2+\alpha^{2})r^{2}-3b^{2}\big)}{(\alpha^{2}+2)\pi L^{2}r_{+}}\big(1-\frac{b^{2}}{r_{+}^{2}}\big)^{\frac{\alpha^{2}-4}{2(2+\alpha^{2})}}\\
\label{ExpreTeam}
&&\quad\quad+\frac{1}{2\pi r_{+}}(1-\frac{b^{2}}{r_{+}^{2}})^{\frac{4-\alpha^{2}}{2(2+\alpha^{2})}}
\eea
in which we have used $A(r_+)=0$ to eliminate the parameter $c$, namely
\bea{}
\label{RepCWithHori}
c=(1+(1-\frac{b^{2}}{r_{+}^{2}})^{\frac{2\alpha^{2}-2}{2+\alpha^{2}}}\frac{r_{+}^{2}}{L^{2}})^{1/2}r_{+}
\eea
According to area laws, the entropy of system is
\bea{}
\label{BHentropy}
S=\frac{2\pi r_{+}^{3}\Omega_{3}}{\kappa_{5}^{2}}(1-\frac{b^{2}}{r_{+}^{2}})^{\frac{3\alpha^{2}}{2(2+\alpha^{2})}}
\eea
From the Gauss law, the electric charge is
\bea{}
\label{BHECharge}
Q=\frac{br_{+}\Omega_{3}}{4\pi}\sqrt{\frac{6}{2+\alpha^{2}}}\sqrt{1+\frac{r_{+}^{2}}{L^{2}}(1-\frac{b^{2}}{r_{+}^{2}})^{\frac{2\alpha^{2}-2}{2+\alpha^{2}}}}
\eea
in which we also use equality $A(r_+)=0$ to eliminate parameter $c$. Finally, the chemical potentials can be caculated as
\bea{}
U=A_{t}\vert_{r\to\infty}-A_{t}\vert_{r\to r_{+}}=-\frac{4\pi q}{\kappa_{5}^{2}r_{+}^{2}}
\eea
where the definition of $A_{t}$ is
\bea{}
A_{t}=-\int dr\frac{8\pi}{\kappa^{2}}F_{tr}=-\int dr\frac{8\pi q}{\kappa^{2}r^{3}}=\frac{8\pi}{2\kappa^{2}}\frac{q}{r^{2}} 
\eea
Replace the parameter $c$ in \eqref{FiExOf} with the equality \eqref{RepCWithHori}, we give
\bea{}
M=\frac{3\Omega_{3}r_{+}^{2}}{2\kappa_{5}^{2}}\bigg(1+(1-\frac{b^{2}}{r_{+}^{2}})^{\frac{2\alpha^{2}-2}{2+\alpha^{2}}}\frac{r_{+}^{2}}{L^{2}}+(\frac{2-\alpha^{2}}{2+\alpha^{2}})\frac{b^{2}}{r_{+}^{2}}\bigg)
\eea
With thermodynamic quantities given above,  we can ensure the rightness of black hole mass \eqref{FiExOf} by cheking that the first law of black hole thermodynamics is holden. According to the link rule of differentiation \cite{Sheykhi:2009pf,Li:2017kkj}, we obtain
\bea{}
\label{PMPS}
&&\hspace{-15mm}\bigg(\frac{\partial M}{\partial S}\bigg)_{Q}=\bigg(\frac{\partial M}{\partial r_{+}}\frac{\partial r_{+}}{\partial b}+\frac{\partial M}{\partial b}\bigg)_{Q}\bigg/\bigg(\frac{\partial S}{\partial r_{+}}\frac{\partial r_{+}}{\partial b}+\frac{\partial S}{\partial b}\bigg)_{Q}\\
\label{MidPMPS}
&&\hspace{-15mm}\bigg(\frac{\partial r_{+}}{\partial b}\bigg)_{Q}=-\bigg(\frac{\partial Q}{\partial b}\bigg)\bigg/\bigg(\frac{\partial Q}{\partial r_{+}}\bigg)\\
\label{PMPQ}
&&\hspace{-15mm}\bigg(\frac{\partial M}{\partial Q}\bigg)_{S}=\bigg(\frac{\partial M}{\partial r_{+}}\frac{\partial r_{+}}{\partial b}+\frac{\partial M}{\partial b}\bigg)_{S}\bigg/\bigg(\frac{\partial Q}{\partial r_{+}}\frac{\partial r_{+}}{\partial b}+\frac{\partial Q}{\partial b}\bigg)_{S}\\
\label{MidPMPQ}
&&\hspace{-15mm}\bigg(\frac{\partial r_{+}}{\partial b}\bigg)_{S}=-\bigg(\frac{\partial S}{\partial b}\bigg)\bigg/\bigg(\frac{\partial S}{\partial r_{+}}\bigg)
\eea
note that the extra minus sign in \eqref{MidPMPS} and \eqref{MidPMPQ} are originated from the derivative of implicit functions $Q(b,r_+)=Q_0, S(b,r_+)=S_0$. After some tedious but straightforward caculations, it's easy to check that
\bea{}
\bigg( \frac{\partial M}{\partial S} \bigg)_Q =T, \quad \quad  \bigg( \frac{\partial M}{\partial Q} \bigg)_S =U
\eea
thus the above thermodynamics quantities indeed satisfy the first law of black hole thermodynamics,
\bea{}
dM=TdS+UdQ
\eea

\section{Brane's motion in spacetime of black hole \label{BranMotion}}

\subsection{Motion of brane in a static bulk with one extra dimension} 

We assume $\mathcal{M}$ is a 5-dimensional manifold containing a brane $\Sigma$ with two sides (denoted by $\Sigma_\pm$ respectively), which splits M into two parts, i.e. $\mathcal{M}_{\pm}$, with a $Z_2$ symmetry. For describing the bulk-brane system clearly, we need to introduce some physical quantities before giving formula expression. We use $X^A$ to denote bulk coordinates and $x^\mu$ to denote internal coordinates of the brane world sheet. If we consider the motion of brane in bulk with one extra dimension, one can then embed brane into bulk with trajectory $X^A(x^\mu)$ (see Table.\ref{EmbedBrane} for more detail), while construct the vielbein as $e^A _\mu =\frac{\partial X^A}{\partial x^\mu}$.
\begin{table}[!ht]
	\begin{center}
		\begin{tabular}{|c|c|c|c|}
			\hline
			Bulk & $t$ & $r$ & $\vec{X}_3$ \\
			\hline
			brane & $t_{b}$ & $\times$ & $\vec{x}_3$ \\
			\hline
			embeded brane & $t(t_b)$ & $r(t_b)$ & $\vec{x}_3=\vec{X}_3$\\
			\hline
		\end{tabular}
		\caption{This table shows how embed brane into bulk in each component of coordinates.}
		\label{EmbedBrane}
	\end{center}
\end{table}			
According to the static gauge, the $t_b$ could be chosen as proper time $\tau$ of bulk coordinates. We will set $t_b=\tau$ in the following contents. 

We consider a static bulk spacetime with the ansatz \eqref{Ge5dBHAnsa}.
The velocity of the brane could be written as $u^M=(\dot{T}(\tau),\dot{r}(\tau),0,0,0)$, normalization $u^M u_M=-1$ gives $\dot{T}(\tau)=\sqrt{(1+B\dot{r}^{2})/A}$. Let $n_M$ be the unit normal pointint into $M_\pm$, according to orthogonal condition $u^M n_M =0, n^M n_M=1$, we obtain $n_{M}=(\sqrt{AB}\dot{r},-\sqrt{B(1+B\dot{r}^{2})},0,0,0)$. The induced metric on $\Sigma_\pm$ is defined by $h_{\mu\nu}=e^M _\mu e^N _\mu g_{AB}$, explicitly 
\begin{align}
\label{induceBrane}
ds^2=-d\tau^2 +R\big( r(\tau) \big)^2 d\Omega ^2 _3
\end{align}

The physics of bulk-brane system could be described by action
\bea{}
\label{BhBraneAct}
&&S=S_{EMD}+S_{brane}\\
\label{BraneAct}
&&S_{brane}=\int dx^{4}\sqrt{-h}\big\{\frac{\mathcal{K}}{\kappa_{5}^{2}}+\lambda(\phi)+\mathcal{L}_{matter}\big\}
\eea
where $\lambda$ is a undetermined function of $\phi$, which represents the effective brane tension. And the $\mathcal{K}$ is the trace of extrinsic curvature tensor \cite{Gibbons:1976ue}. Varying $\eqref{BhBraneAct}$ with respect to metric tensor. Besides getting a standard Einstein equation in bulk spacetime, we also obtain a Israel junction condition on brane \cite{Israel:1966rt,Chamblin:1999ya},
\bea{}
\label{OriIsrael}
\big\{\mathcal{K}_{\mu\nu}-\mathcal{K}h_{\mu\nu}\big\}\vert_{\Sigma_{+}}-\big\{\mathcal{K}_{\mu\nu}-\mathcal{K}h_{\mu\nu}\big\}\vert_{\Sigma_{-}}=\kappa_{5}^{2}S_{\mu\nu}
\eea
where $S_{\mu \nu}$ is the energy-momentum tensor of matters which are confined on brane, and the $\mathcal{K}_{\mu \nu}$ is defined as
\bea{}
\label{ExtrinCurva}
\mathcal{K}_{\mu \nu}=\frac{1}{2} e^M_{\mu} e^N_{\nu} \big( \nabla_M n_N +\nabla_N n_M  \big)
\eea
We will involve a $Z_2$ symmetry in two sides of brane, namely $\big\{\mathcal{K}_{\mu\nu}\big\}\vert_{\Sigma_{-}}=-\big\{\mathcal{K}_{\mu\nu}\big\}\vert_{\Sigma_{+}}$. In this way, the junction condition $\eqref{OriIsrael}$ could be further simplified as
\bea{}
\label{z2juncon}
\mathcal{K}_{\mu\nu}-\mathcal{K}h_{\mu\nu}=\frac{\kappa_{5}^2}{2} \mathcal{S}_{\mu\nu}
\eea
for clearness in expression, we omit the subscript $\Sigma_{+}$. Alternatively, \eqref{z2juncon} could also be written as
\bea{}
\label{alterz2juncon}
\mathcal{K}_{\mu\nu}=-\frac{\kappa^2_5}{2} \big( S_{\mu\nu}-\frac{1}{3}Sh_{\mu\nu} \big)
\eea
For general ansatz of 5-dimensional static black hole \eqref{Ge5dBHAnsa}, the nonvanishing components of $\mathcal{K}_{\mu \nu}$ are given by
\bea{}
\label{ExtKij}
&&\mathcal{K}_{ij}=-\frac{R^{\prime}}{R}\frac{\sqrt{1+B\dot{r}^{2}}}{\sqrt{B}}h_{ij}\\
\label{ExtKtautau}
&&\mathcal{K}_{\tau\tau}=\frac{\sqrt{\big(1+B\dot{r}^{2}\big)}}{2A\sqrt{B}}\bigg(A^{\prime}+\big(AB^{\prime}-A^{\prime}B\big)\dot{r}^{2}\bigg)
\eea
in which $\dot{r}=dr(\tau)/d\tau$, and we will also use dot to denote the derivative with respect to $\tau$ in following contents. Varying \eqref{BhBraneAct} with respect to dilaton field. Besides giving the equation of motions \eqref{eqphi} in bulk, we also obtain a boundary condition of dilaton field on brane,
\bea{}
\label{JunScalar}
\frac{4}{3\kappa_{5}^{2}}n^{M}\partial_{M}\phi=\frac{\partial\lambda}{\partial\phi}
\eea 
Expanding the \eqref{JunScalar} yields
\bea{}
\label{DetaJunScalar}
-\frac{\sqrt{(1+B\dot{r}^{2})}}{\sqrt{B}}\partial_{r}\phi=\frac{3\kappa_{5}^{2}}{4}\frac{\partial\lambda}{\partial\phi}
\eea
Besides, if we set the energy-momentum tensor of matters on brane as simple ideal fluid types, namely
\bea{}
\label{Exoidealfluid}
\mathcal{S}^\mu _\nu=Diag\{-\rho,P,P,P \}
\eea
then by combining $\eqref{Exoidealfluid}$ with $\eqref{BraneAct}$, we will obtain the two independent components of the Israel junction condition as
\bea{}
\nonumber
&&\hspace{-19mm}\frac{\sqrt{(1+B\dot{r}^{2})}}{2A\sqrt{B}}\big(A^{\prime}+(AB^{\prime}-A^{\prime}B)\dot{r}^{2}\big)+\frac{2R^{\prime}\sqrt{1+B\dot{r}^{2}}}{R\sqrt{B}}\\
 \label{ExpIsrael1}
&&\quad \quad \quad \quad \quad \quad \quad \quad \quad \quad=\frac{\kappa_{5}^{2}}{2}(\lambda+P)\\
 \nonumber
 \\
 \label{ExpIsrael2}
&&\quad\frac{R^{\prime}}{R}\frac{\sqrt{1+B\dot{r}^{2}}}{\sqrt{B}}=\frac{\kappa_{5}^{2}}{6}(\lambda-\rho)
\eea

\subsection{Effective Einstein field equations on  brane}

\begin{figure}[ht]
	\begin{center}
		\includegraphics[scale=0.29]{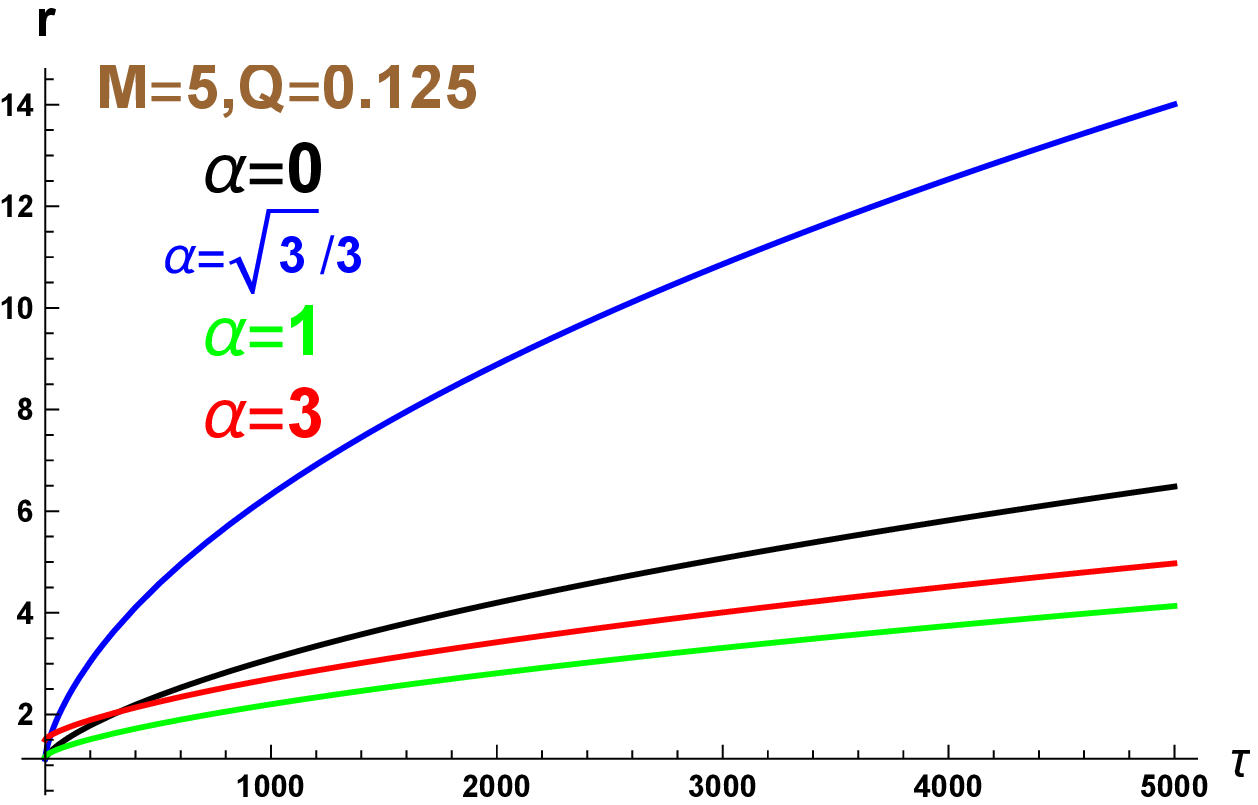}
		\includegraphics[scale=0.29]{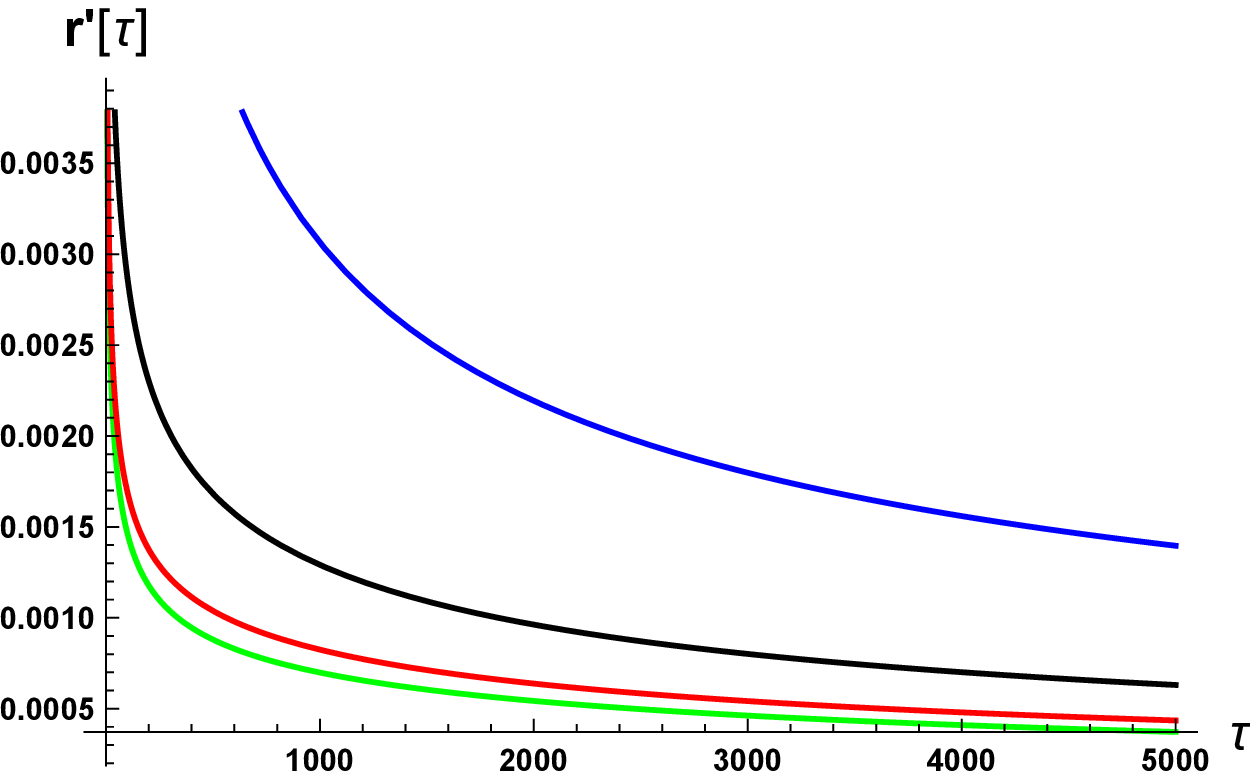}\\
		\includegraphics[scale=0.29]{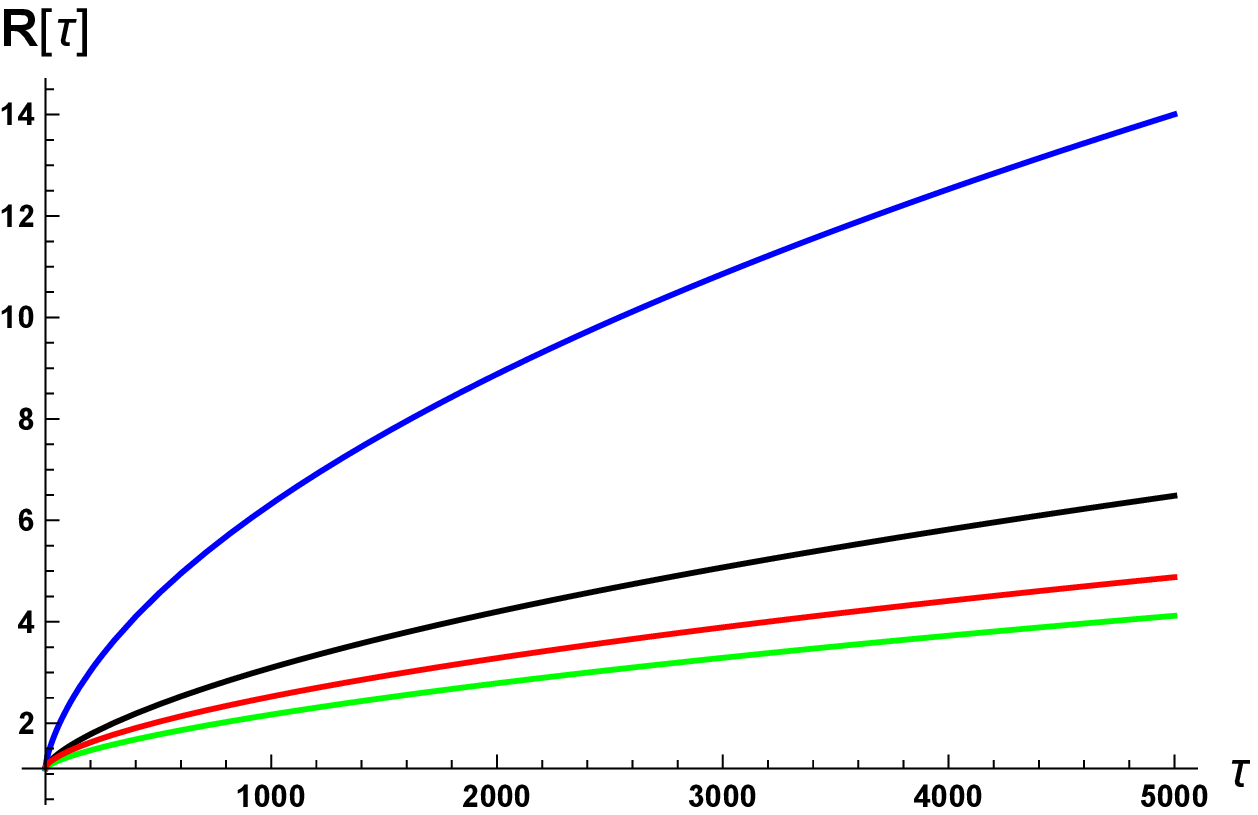}
		\includegraphics[scale=0.29]{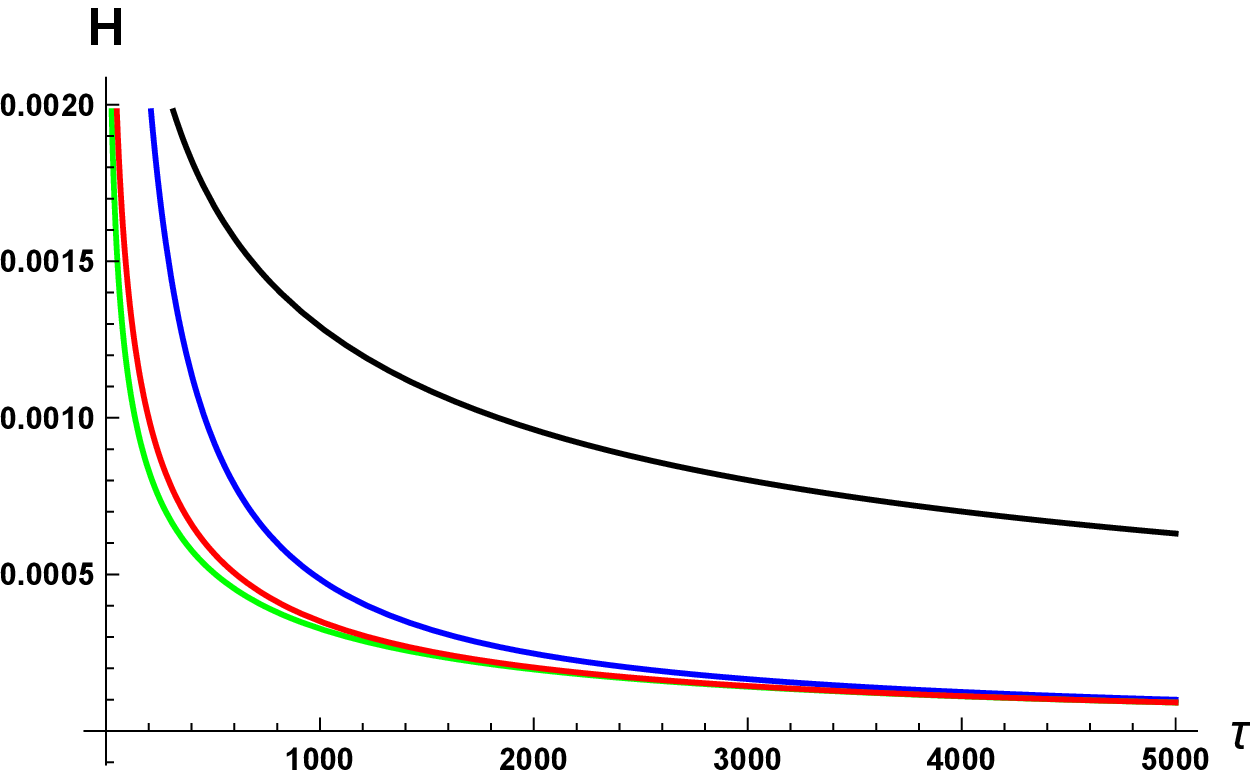}
		\caption{(color online). The upper part displays the location $r(\tau)$ and velocity $\frac{dr(\tau)}{d\tau}$ of brane in bulk coordinates as the increase of proper time. The bottom part plots the evolution of effective scale factor $R(r(\tau))$ and Hubble constant $H=\frac{dR/d\tau}{R}$ for the observer confined on brane universe. The red, green, blue, and black curves correspond the $\alpha=3,\alpha=1,\alpha=\sqrt{3}/3,\alpha=0$ respectively, meanwhile all of these curves correspond the same black hole mass and charge. Both the AdS radius and $\kappa$ are set to be 1.}
		\label{Solrtau}
	\end{center}
\end{figure}

As shown in \cite{Shiromizu:1999wj}, the effective Einstein field equation on brane is
\bea{}
\nonumber
&&\hspace{-4mm}\frac{3}{2}\bigg(\mathcal{R}_{\mu\nu}-\frac{1}{6}h_{\mu\nu}\mathcal{R}\bigg)+\frac{3}{2}E_{\mu\nu}-\frac{3}{2}\mathcal{K}\mathcal{K}_{\mu\nu}+\frac{3}{2}\mathcal{K}_{\mu\rho}\mathcal{K}_{\nu}^{\rho}+\frac{1}{4}h_{\mu\nu}\mathcal{K}^{2}\\
\label{effEinOnbrane}
&&\hspace{-4mm}-\frac{1}{4}h_{\mu\nu}\mathcal{K}_{\alpha\beta}\mathcal{K}^{\alpha\beta}=\kappa_{5}^{2}\bigg(\mathcal{T}_{MN}h_{\mu}^{M}h_{\nu}^{N}-\frac{1}{4}h_{\mu\nu}\mathcal{T}\bigg)
\eea
in which $R_{\mu\nu}$ and $R$ is the Ricci tensor and Ricci scalar of the induced metric $h_{\mu \nu}$, while the $E_{\mu \nu}$ is defined as
\bea{}
\label{tidaltensor}
E_{\mu \nu}=\mathcal{C}_{MNAB} n^M  n^A e^N _\mu e^B _\nu
\eea
where the $\mathcal{C}$ is the Weyl curvature tensor,
\bea{}
\nonumber
\mathcal{C}_{MNAB}=R_{MNAB}-\frac{2}{3}(g_{MA}R_{BN}-g_{NA}R_{BM})+\frac{1}{6}g_{MA}g_{BN}R
\eea
substitute $\eqref{induceBrane},\eqref{z2juncon}$ into $\eqref{effEinOnbrane}$ and expand it explicitly, we give two equations as following
\bea{}
\nonumber
&&\frac{3A^{\prime}}{2AB}\frac{R^{\prime}}{R}+\frac{3A^{\prime}}{2A}\frac{R^{\prime}}{R}\dot{r}^{2}+\frac{3B^{\prime}}{2B}\frac{R^{\prime}}{R}\dot{r}^{2}-\frac{3}{2B}\frac{(R^{\prime})^{2}}{R^{2}}-\frac{3}{2}\frac{(R^{\prime})^{2}}{R^{2}}\\
\label{EffEinOnbrane1}
&&+3\frac{R^{\prime}}{R}\ddot{r}=\frac{\kappa_{5}^{4}\lambda}{4}P+\frac{\kappa_{5}^{4}\lambda^{2}}{24}-\frac{\kappa_{5}^{4}}{4}P\rho+\frac{\kappa_{5}^{4}\lambda}{6}\rho-\frac{5\kappa_{5}^{4}}{24}\rho^{2}\\
\label{EffEinOnbrane2}
&&\frac{3}{2B}\frac{(R^{\prime})^{2}}{R^{2}}\big(1+B\dot{r}^{2}\big)=\frac{\kappa_{5}^{4}\lambda^{2}}{24}-\frac{\kappa_{5}^{4}\lambda}{12}\rho+\frac{\kappa_{5}^{4}}{24}\rho^{2}
\eea
Note that the $R$ represents the solution \eqref{solR}, which is the function of $r$ in metric \eqref{Metric} rather than the Ricci scalar. It's easy to see that the $\eqref{EffEinOnbrane2}$ is same with $\eqref{ExpIsrael2}$. Thus the bulk-brane system is described by 4 independent equations $\eqref{DetaJunScalar},\eqref{ExpIsrael1},\eqref{ExpIsrael2},\eqref{EffEinOnbrane1}$. When combining with the Einstein field equation in bulk $\eqref{Eintt}$-$\eqref{Einxx}$  and Israel junction condition $\eqref{ExpIsrael1}$-$\eqref{ExpIsrael2}$, the $\eqref{EffEinOnbrane1}$ could be further simplified as
\bea{}
\label{DeterEvort}
\ddot{r}+\frac{A^{\prime}}{2A}\dot{r}^{2}+\frac{A^{\prime}B}{2A}\dot{r}^{4}-\frac{1}{2}B^{\prime}\dot{r}^{4}=0
\eea
After substituting $\eqref{solA}$, $\eqref{solB}$ into $\eqref{DeterEvort}$ and solving this differential equation numerically at fixed parameters $M$ and $Q$ with different $\alpha$. We display the variation of brane's position $r(\tau)$ and corresponding velocity $\dot{r}(\tau)$ in top-left and top-right panel of Fig.\ref{Solrtau}. Note that for the observer confined on the brane universe, the scale factor is $R(r(\tau))$ according to the induced metric $\eqref{induceBrane}$, thus we also show the variation trend of $R(r(\tau))$ and corresponding Hubble constant $H=\frac{\dot{R}}{R}=\frac{R^\prime (r(\tau))}{R}\dot{r}$ in Fig.\ref{Solrtau}.

\begin{figure}[ht]
	\begin{center}
		\includegraphics[scale=0.43]{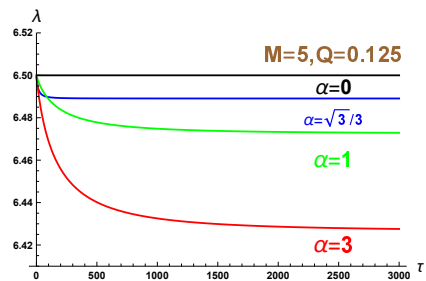}
		\caption{(color online). According to the equation \eqref{InteFofLamtau}, we plot the evolution of effective brane tension $\lambda$ as the function of proper time $\tau$. The initial brane's tension is set as $\lambda_0=6.5$. Both the AdS radius and $\kappa$ are set to be 1.}
		\label{SolLamtau}
	\end{center}
\end{figure}

From the $\eqref{DetaJunScalar}$, we obtain the mathmatical expression of brane tension $\lambda$ as the integrate function of $\tau$,
\bea{}
\label{InteFofLamtau}
&&\lambda=\lambda_{0}-\frac{4}{3\kappa_{5}^{2}}\int\frac{\sqrt{(1+B\dot{r}^{2})}}{\sqrt{B}}(\phi^{\prime})^{2}\dot{r}d\tau
\eea
Basing on $\eqref{InteFofLamtau}$, we give the variation trend of $\lambda$ with respect to $\tau$ in Fig.\ref{SolLamtau}.
Combining the equations \eqref{ExpIsrael1} and \eqref{ExpIsrael2} with the numerical solutions $r(\tau)$ and $\lambda(\tau)$, we can also plot the  evolution of energy density $\rho$ and pressure $P$ of the matters confined on brane as follows
\begin{figure}[ht]
	\begin{center}
		\includegraphics[scale=0.29]{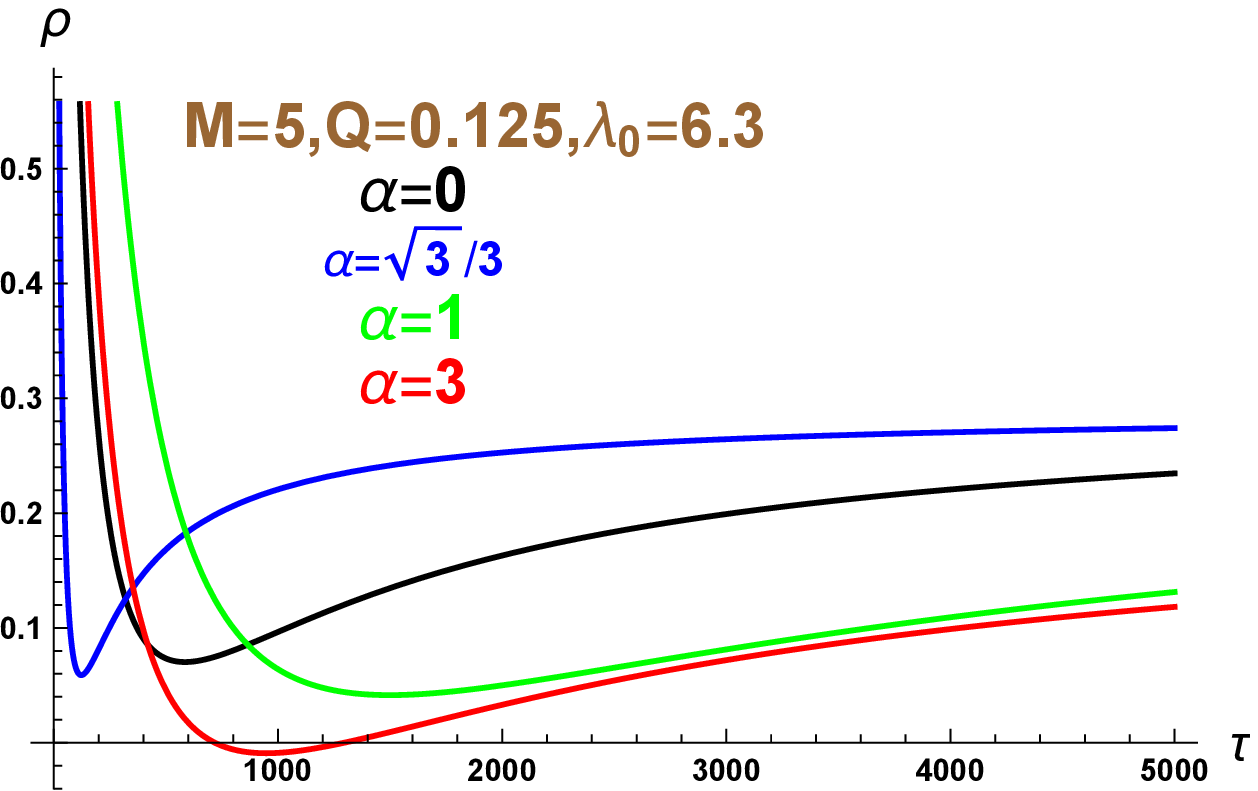}
		\includegraphics[scale=0.29]{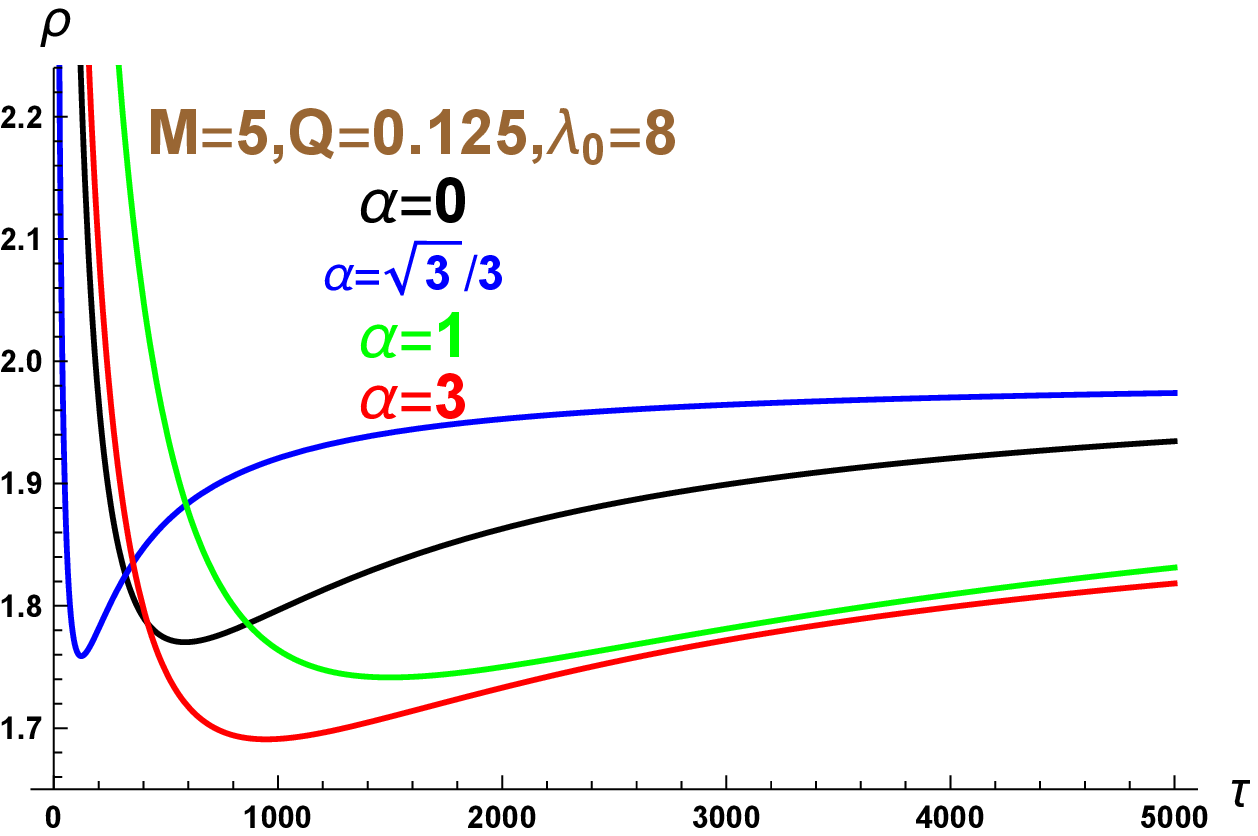}\\
		\includegraphics[scale=0.29]{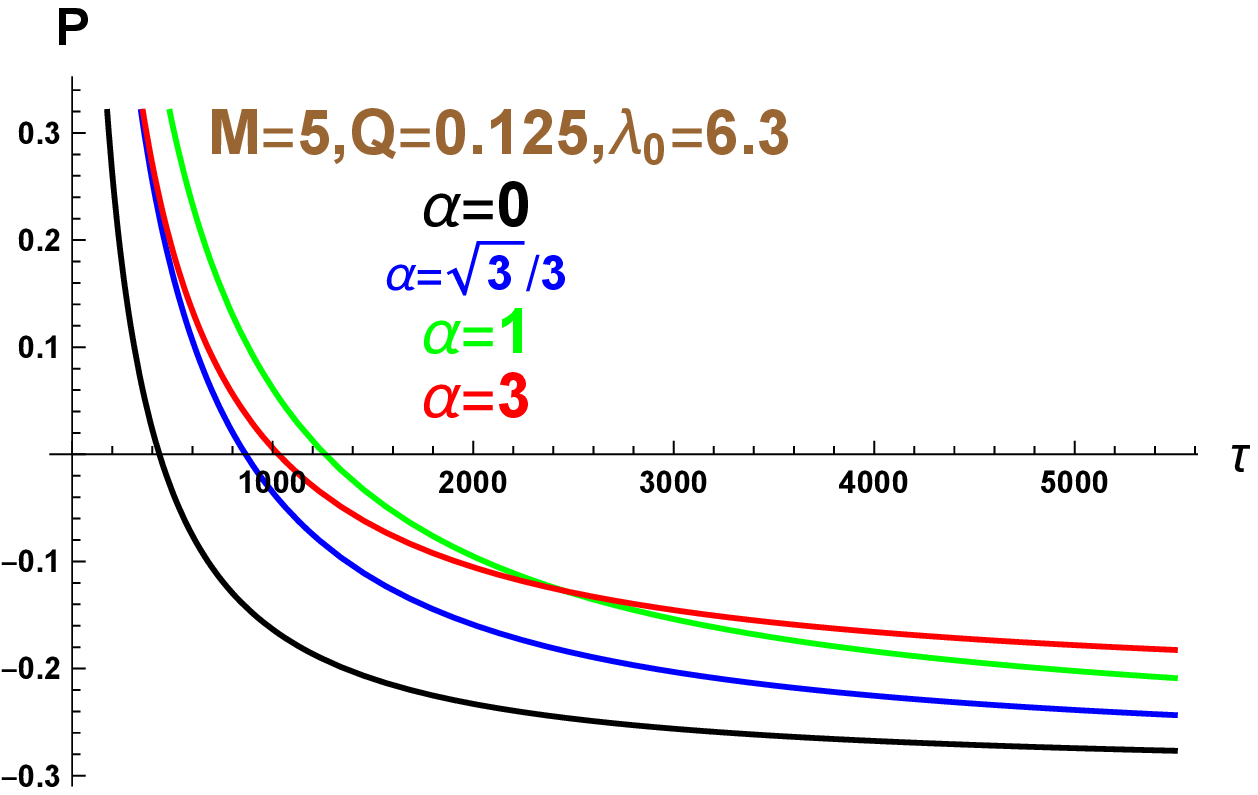}
		\includegraphics[scale=0.29]{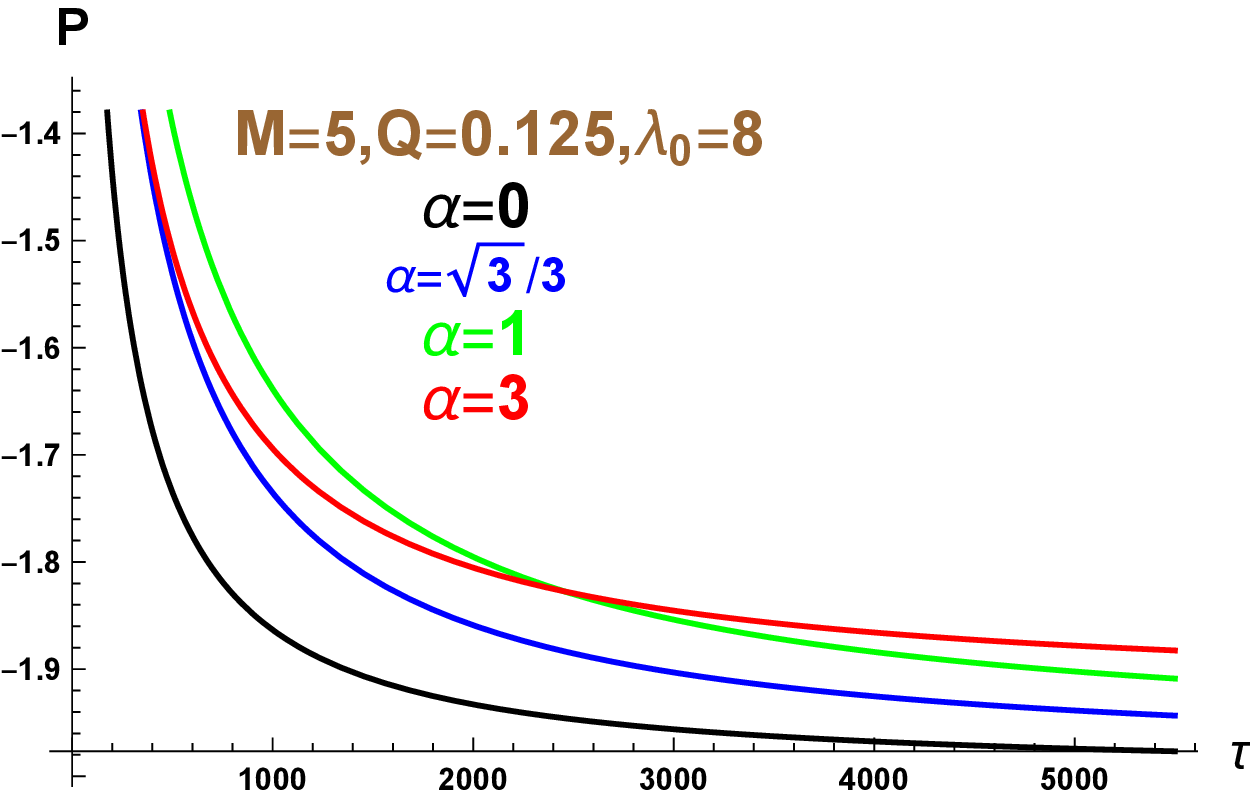}
		\caption{(color online). The evolution of energy density $\rho$ and preessure $P$ for the matters confined on brane, at fixed black hole mass $M$ and charge $Q$ with different dilaton coupling constant $\alpha$ and initial brane tension $\lambda_0$. Both the AdS radius and $\kappa$ are set to be 1.}
		\label{RhotPt}
	\end{center}
\end{figure}

\section{Holographic effects on  brane \label{BranHolo}}

\subsection{A renormalization procedure inspired by AdS/CFT}

As indicated in \cite{Witten:1998qj}, the energy, entropy and temperature of the CFT on boundary spacetime at high temperatures could be identified with the mass, Bekenstein-Hawking entropy, and Hawking temperature of AdS black hole. Thus, the matters on brane will have the CFT's feature when brane approaches the AdS boundary. In bulk spacetime, the total energy is just the black hole mass\eqref{FiExOf}, namely $E=M$. In case of large $r(\tau)$, the matter's energy on brane is given by
\bea{}
\label{EonBatLr}
E=M\dot{t} \approx ~\frac{3\Omega_{3}}{2\kappa_{5}^{2}}(c^{2}+\frac{2-\alpha^{2}}{2+\alpha^{2}}b^{2})~ \frac{L}{r}
\eea
Note that the $E=M$ is obtained by using the bulk's time coordinate $t$. Nevertheless an observer on brane measures the total energy by using the time coordinate $\tau$, thus we need to scale $E$ by $\dot{t}$. And the energy density is given by
\bea{}
\label{rhoCFT}
\rho=\frac{E}{V}=\frac{3}{2\kappa_{5}^{2}}(c^{2}+\frac{2-\alpha^{2}}{2+\alpha^{2}}b^{2})\frac{L}{r^{4}}
\eea 
where the $V=(R(r(\tau)))^3 \Omega_3$ is just the spatial volume of the brane, from \eqref{solPhi} we know that $R(r(\tau)) = r(\tau)$ in the large $r(\tau)$ limit. Meanwhile, we use the $r$ to denote $r(\tau)$ in \eqref{EonBatLr}, \eqref{rhoCFT} for convenience, and this convention will still be used in this and next section. Namely, without pointing out explicitly, the $r$ will always represent the $r(\tau)$ in Sec\ref{BranHolo}. According to the first law of thermodynamics,
\bea{}
\label{FirstTher} 
TdS+UdQ=dE+PdV
\eea
the pressure is obtained as
\bea{}
\label{pressCFT}
P=-\frac{\partial E}{\partial V}=\frac{L}{2\kappa_{5}^{2}r^{4}}\big(c^{2}+\frac{2-\alpha^{2}}{2+\alpha^{2}}b^{2}\big)
\eea
It's easy to check that $P=\frac{1}{3}\rho$. However, from the Fig.\ref{RhotPt}, we observe that $P=-\rho$ in region of large $r$.  

Actually, from the scenario of brane-bulk geometry, there exists a motional brane which has cut off the bulk spacetime. Translate the scenario of geometry into the word of quantum field theory, it means that the matter field confined on brane is a symmetry broken effective field theory of CFT on AdS boundary at lower energies. Namely, in the case of $r(\tau) \ll \infty$, the Israel junction could be re-expressed as
\bea{}
\mathcal{K}_{\mu \nu} -\mathcal{K}h_{\mu \nu}=\frac{\kappa^2_5}{2}\big(\langle S^{CFT}_{\mu \nu}\rangle+\langle S^{IR}_{\mu \nu} \rangle+\lambda h_{\mu\nu} \big)
\eea 
When the brane approaches the AdS boundary, namrly $r(\tau)\to \infty$, we need add a counterterms to subtract the infrared(IR) divergences and restore the conformal symmetry, finally the modified Israel junction condition becomes
\bea{}
\label{ModiIsrael}
\mathcal{K}_{\mu\nu}-\mathcal{K}h_{\mu\nu}-\frac{\kappa_{5}^{2}}{2}S_{\mu\nu}^{ct}=\frac{\kappa_{5}^{2}}{2}\big(\langle S_{\mu\nu}^{CFT}\rangle+\lambda h_{\mu\nu}\big)
\eea 
As considered in \cite{deHaro:2000wj, Gubser:1999vj, Anchordoqui}, the action of countermterms could be involved as the Ricci scalar $R$ and higher order curvature terms respect to brane's induced metric $h_{\mu\nu}$. To be consistent with \eqref{CounterAction}, we choose the counterterm as the following ansatz
\bea{}
\label{ctBraneLr}
&&\hspace{-8mm}S_{ct}=-\frac{1}{\kappa_{5}^{2}}\int_{\Sigma}d^{4}x\sqrt{-h}\big\{ b_{1}l_{eff}\mathcal{R}+b_{2}l_{eff}^{3}\mathcal{R}^{2} \big\}
\eea
where $l_{eff}(\phi)=\sqrt{-\frac{3(3+1)}{V(\phi)}}$. Similarly, we have ignored the $\mathcal{R}^{\mu\nu} \mathcal{R}_{\mu\nu}$ term for simplification in caculation, without loss of generality. From $\eqref{ctBraneLr}$, the corresponding $S^{ct}_{\mu\nu}$ could be caculated as 
\bea{}
\nonumber
&&S^{ct}_{\mu\nu}=-\frac{2c_{1}l_{eff}}{\kappa_{5}^{2}}\bigg(\mathcal{R}_{\mu\nu}-\frac{1}{2}\mathcal{R}h_{\mu\nu}\bigg)+\frac{c_{2}}{\kappa_{5}^{2}}l_{eff}^{3}\bigg\{h_{\mu\nu}\mathcal{R}^{2}\\
\label{ctTenBrane}
&&\quad\quad-4\mathcal{R}\mathcal{R}_{\mu\nu}+4\nabla_{\nu}\nabla_{\mu}\mathcal{R}-4h_{\nu\mu}\nabla_{\alpha}\nabla^{\alpha}\mathcal{R}\bigg\} 
\eea
Meanwhile, we assume the $\lambda(\phi)$ has the following mathmatical form in case of large $r(\tau)$
\bea{}
\label{AsymLam}
\lambda(\phi)=\frac{1}{\kappa_{5}^{2}}\frac{b_{0}}{l_{eff}(\phi)}\big(1+\frac{b_{\alpha}}{b_{0}}\phi^{2}\big)
\eea
 the coefficients $b_0, b_\alpha, b_1, b_2$ will be decided later.
 
 Substitute $\eqref{AsymLam}$ and $\eqref{ctTenBrane}$ into the modified Israel junction condition $\eqref{ModiIsrael}$, and expand $\rho, P$ in the case of large $r(\tau)$, we obtain
 \bea{}
 \nonumber
 &&\hspace{-3mm}\kappa^2_5\rho=\frac{(b_{0}-6)}{L}-\frac{3L(1-2b_{1})}{r^{2}}+\frac{3}{4L(2+\alpha^{2})^{2}r^{4}}\big\{\\
\nonumber
&&\hspace{-3mm}\quad\quad~\alpha^{2}b^{4}(3b_{\alpha}-8)+(2+\alpha^{2})^{2}L^{2}\big(4c^{2}+(1+48b_{2})L^{2}\big) \\
 \label{HoloModRho}
&&\hspace{-3mm}\quad\quad~+8(2+\alpha^{2})\big(1+(b_{1}-1)\alpha^{2}\big)b^{2}L^{2}\big\}+O(\frac{1}{r^{6}})\\
\nonumber
\\
\nonumber
&&\hspace{-3mm} \kappa^2_5P=\frac{(b_{0}-6)}{L}+\frac{(1-2b_{1})L}{r^{2}}+\frac{1}{4L^{2}(2+\alpha^{2})^{2}r^{4}}\big\{\\
\nonumber
&&\hspace{-3mm}\quad\quad~3\alpha^{2}b^{4}L(8-3b_{\alpha})+(2+\alpha^{2})^{2}L^{3}(4c^{2}+(1+48b_{2})L^{2})\\
\label{HoloModP}
&&\hspace{-3mm}\quad\quad~-8(2+\alpha^{2})(b_{1}\alpha^{2}-1)b^{2}L^{3}\big\}+O(\frac{1}{r^{6}})
 \eea 
Note that the equation $\eqref{DeterEvort}$ is invariant, although the Israel junction condition is changed from $\eqref{OriIsrael}$ to $\eqref{ModiIsrael}$. So, the evolution of $r(\tau)$ is still controlled by the equation $\eqref{DeterEvort}$. It's easy to observe from fig.\ref{Solrtau} that the $\dot{r}$ and higher order derivatives will approaches 0 when $r(\tau)\to \infty$, thus we have ignored all $O(\frac{1}{r^2}), O(\frac{1}{r^4})$ terms which include the derivatives of $r(\tau)$. Finally, by comparing the \eqref{HoloModRho} and \eqref{HoloModP} with the expected results \eqref{rhoCFT} and \eqref{pressCFT}, we can give these undetermined coefficients as $b_0=6,~b_1=\frac{1}{2},~b_\alpha=\frac{8}{3},~b_2=-\frac{1}{48}$.

\begin{figure}[ht]
	\begin{center}
		\includegraphics[scale=0.58]{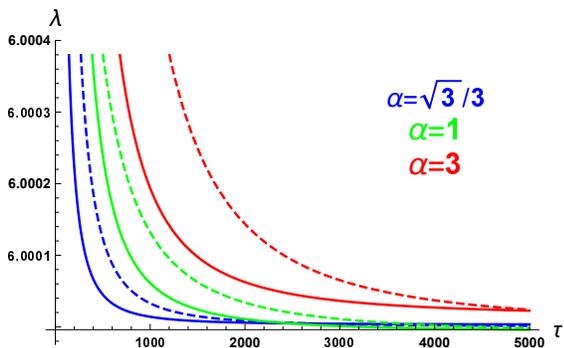}
		\caption{(color online). The dotted curves correspond to the approximate solution of effective brane tension $\lambda$ given by \eqref{AsymLam} which is valid in large $r(\tau)$ case (as shown in Fig.\ref{Solrtau}, the $r(\tau)$ is the monotonically increasing function of $\tau$, and hence the large $r(\tau)$ value also means the large $\tau$ value). The solid curves correspond to the  numerically exact solution of $\lambda$ given in \eqref{InteFofLamtau}. All these curves have the same parameter value $M=5,Q=0.125,L=1,\kappa=1$. The value of $\lambda_0$ set for the solid-red, solid-green, solid-blue curves are 6.0737, 6.02757, 6.01093 respectively.}
		\label{efflamlati}
	\end{center}
\end{figure}

Substitute $b_0=6,~b_\alpha=\frac{8}{3}$ into the \eqref{AsymLam}, we obtain a analytical expression for the effective brane tension as function of dilaton field
\bea{}
\label{AnaLamPhiLaTi}
\lambda(\phi)=\frac{1}{\kappa_{5}^{2}}\frac{6}{l_{eff}(\phi)}\big(1+\frac{4}{9}\phi^{2}\big)
\eea
Next, we check that the \eqref{AnaLamPhiLaTi} is indeed valid in large $r(\tau)$ case. Combine the \eqref{AnaLamPhiLaTi} with equation \eqref{eqphi} and the numerical solution of $r(\tau)$, we can plot the evolution of $\lambda$ with respect to the $\tau$ as shown by the dotted curves in Fig.\ref{efflamlati}. Meanwhile, the \eqref{InteFofLamtau} is also displayed in Fig.\ref{efflamlati} by the solid curves. It is obviously shown that by adjusting the parameter $\lambda_0$ appropriately, \eqref{AnaLamPhiLaTi} is consistent with the \eqref{InteFofLamtau} at late time. Besides the validity of \eqref{AnaLamPhiLaTi} in large $r(\tau)$ case, Fig.\ref{efflamlati} also reveals a physical fact that a fixed $\lambda_0$ is implied by the AdS/CFT correspondence.

\subsection{Cardy-Verlinde Formula}

The entropy of (1+1)-dimensional CFT system could be given by a very famous formula, namely the so-called Cardy formula \cite{Cardy:1986ie}. By using the combination of AdS/CFT correspondence and thermodynamics of AdS-Schwarzschild black holes, \cite{Verlinde:2000wg, Savonije:2001nd} has generalized the Cardy formula to the CFT system which lives in arbitrary-dimensional spacetime, this generalized Cardy formula is called as the Cardy-Verlinde formula. After that, the Cardy-Verlinde formula have been checked in various black holes with AdS asymptotics \cite{Klemm:2001db, Cai:2001jc, Birmingham:2001vd, Youm:2001yq, Cai:2001ja, Youm:2001qr, Jing:2002aq, Cai:2002bn, Lee:2008yqa, BravoGaete:2017dso}. In our case, as we have argued in last section, the matters confined on brane will behave as CFT system when brane approaches the AdS boundary. Thus, we naturally expect that the Cardy-Verlinde formula will hold in case of large $r(\tau)$. 

In the limit of $r(\tau)\to\infty$, the temperature and chemical potential of the corresponding CFT on brane could be given as
\bea{}
\nonumber
&&T=T_{H}\dot{t}\approx\frac{L}{2\pi rr_{+}}(1-\frac{b^{2}}{r_{+}^{2}})^{\frac{4-\alpha^{2}}{2(2+\alpha^{2})}}\\
\label{TemaCFT}
&&\quad \quad \quad +\frac{\big((2+\alpha^{2})r_{+}^{2}-3b^{2}\big)}{(\alpha^{2}+2)\pi Lrr_{+}}\big(1-\frac{b^{2}}{r_{+}^{2}}\big)^{\frac{\alpha^{2}-4}{2(2+\alpha^{2})}}\\
\nonumber
\\
\nonumber
&&U=U\dot{t}\approx-\frac{4\sqrt{6}\pi bL}{\sqrt{2+\alpha^{2}}\kappa_{5}^{2}rr_{+}}\\
\label{UCFT}
&&\quad \quad \quad \quad \times\sqrt{1+\frac{r_{+}^{2}}{L^{2}}(1-\frac{b^{2}}{r_{+}^{2}})^{\frac{2\alpha^{2}-2}{2+\alpha^{2}}}}
\eea
The entropy and electric charge are given directly by \eqref{BHentropy} and \eqref{BHECharge}. And we obtain the entropic and electric density as
\bea{}
\label{CFTEntroDen}
&&s=\frac{S}{V}=\frac{2\pi r_{+}^{3}}{\kappa_{5}^{2}r^{3}}(1-\frac{b^{2}}{r_{+}^{2}})^{\frac{3\alpha^{2}}{2(2+\alpha^{2})}}\\
\nonumber
\\
\nonumber
&&\rho_e=\frac{Q}{V}=\frac{\sqrt{6}br_{+}}{4\pi\sqrt{2+\alpha^{2}}r^{3}}\\
\label{CFTElecDen}
&&\quad \quad ~ \times \sqrt{1+\frac{r_{+}^{2}}{L^{2}}(1-\frac{b^{2}}{r_{+}^{2}})^{\frac{2\alpha^{2}-2}{2+\alpha^{2}}}}
\eea
Using the symbol in \cite{Savonije:2001nd}, the $\gamma$ quantity which is relevant with the Casimir energy $E_c$ of CFT system on brane through the relation $E_c=(\frac{\gamma}{r^2})\cdot r^3 \Omega_3$, could be defined as
\bea{}
\nonumber
&&\frac{\gamma}{r^{2}}=3(\rho_{CFT}+P_{CFT}-\vert\rho_{e}U\vert-Ts)-\frac{\gamma_{proper}}{r^{2}}\\
\label{DefiGamma}
&&\quad ~ =3\frac{r_{+}^{2}}{\kappa_{5}^{2}r^{4}}L\bigg(1-\frac{b^{2}}{r_{+}^{2}}\bigg)
\eea
where
\bea{}
\label{GammaProper}
\frac{\gamma_{proper}}{r^{2}}=\frac{r_{+}^{2}}{\kappa_{5}^{2}r^{4}}\bigg(\frac{6}{2+\alpha^{2}}\frac{b^{2}}{r_{+}^{2}}L\bigg)
\eea
the $(\frac{\gamma_{proper}}{r^{2}}) \cdot r^3 \Omega_3$ has the physical meaning of proper internal energy which is viewed as the zero temperature energy of CFT \cite{Cai:2001jc}. Note also that the proper internal energy has no physical effects on the entropy, thus it needs to be subtracted from the total energy when considering the Cardy-Verlinde formula. Combine the physical quantity $\gamma, \gamma_{proper}$ with the energy density $\rho$ in $\eqref{rhoCFT}$, it's easy to find that the entropy density of brane universe in case of $r(\tau) \to \infty$ could be expressed as
\bea{}
\label{CarVerlinFormu}
\big(\frac{3}{2\pi}s\big)^{2}=\gamma \big(2(\rho_{CFT}-\frac{\gamma_{proper}}{r^{2}})-\frac{\gamma}{r^{2}}\big)
\eea 
Compare with the original version of Cardy-Verlinde formula \cite{Savonije:2001nd}, our result $\eqref{CarVerlinFormu}$ has a extra physical quantity $\gamma_{proper}/r^2$. As explained in \cite{Cai:2001jc}, the $\gamma_{proper}/r^2$ is relevant with the proper internal energy which makes the contribution to the free energy of CFT system, but not to the entropy. Thus, the contribution of the proper internal energy must be subtracted when we consider the relationship between the entropy and energy of CFT system on brane. Besides, when $\alpha=0$, the $\eqref{CarVerlinFormu}$ will return the Cardy-Verlinde formula $(3.16)$ in \cite{Cai:2001jc}, which corresponds to the CFT system living on the boundary of AdS-RN black hole spacetime.

\section{Conclusion and Discussion \label{ConAndDis}}

In the background of charged AdS dilaton black hole solved by \cite{Gao:2004tu,Gao:2004tv,Gao:2005xv}, we study the movement of a self-graviting brane and holographic effects as the brane gets close to AdS boundary. Although the movement of brane/wall in this AdS dilaton black hole have been investigated by \cite{Xu:2019abl}, they ignore the self-gravitating effects of brane/wall. Specifically, they assume that there only exists matter field on brane, and the evolution of brane is controlled by the Israel junction condition only. But a more realistic case is that the effective gravitational field should also be localized on brane, thus the evolution of brane is determined by the effective Einstein field equation and Israel junction condition together. In our scenario, we derive the effective Einstein field equation on brane by using the method provided by \cite{Shiromizu:1999wj}, in which they localize the gravitational field by projecting the bulk's Riemann curvature and it's contractions (namely the Ricci tensor and scalar curvature) on brane spacetime. In results of \cite{Xu:2019abl}, they obtain the different evolution modes of brane/wall in various physical parameters, especially for $\omega=\frac{P}{\rho}$ and dilaton coupling constant $\alpha$. However, when taking inclusion of self-graviting effects of brane, we find that the evolution mode of brane is not so sensitive to the value of parameters $\alpha$ and $\omega$. Specifically, as shown in Fig.\ref{Solrtau}, we observe that the evolution of brane only in mode of decelerated expansion whatever the value of $\alpha$, while note that the velocity $\dot{r}$ of brane will tend to be zero as brane gets close to the AdS boundary. On the other hand, from the Fig.\ref{RhotPt}, it is obviously shown that the $\omega=\frac{P}{\rho}$ will approach a fixed vaule $-1$ as time increases.

Inspired by AdS/CFT correspondence, we expect to observe a connection between the brane universe and physical quantities of AdS dilaton black hole when the brane gets close to AdS boundary. Nevertheless, in our scenario of brane-bulk geometry, the AdS spacetime is cut off by a motional brane. And, the matter confined on brane is described by a general quantum field theory without conformal symmetry. Thus, as the brane move close to the boundary of AdS dilaton black hole, a renormalization procedure should be considered to cancel the IR divergences and restore the conformal symmetry. Note that the IR divergences on the gravitational side correspond to the UV divergence on the CFT side. Furthermore, as indicated in \cite{Savonije:2001nd}, when the brane approaches the AdS boundary, AdS/CFT correspondence implies that a radiation dominated FLRW-universe ($P=\frac{1}{3}\rho$) should be given. We add an appropriate surface counterterm to the gravitational action by using the method given in \cite{deHaro:2000wj, Gubser:1999vj, Anchordoqui}, after the inclusion of this counterterm, the correct energy density and pressure (namely $P=\frac{1}{3}\rho$) could be reproduced on brane in limit of $r(\tau)\to \infty$.  It's remarkable that this surface counterterm also plays a important role in caculating the black hole mass. What's more, basing on this holographic renormalization procedure, the evolution of brane's tension as the function of dilaton field could be given analytically in case of large $r(\tau)$. 

By using a holographic method provided in \cite{Verlinde:2000wg}, the temperature, entropy, charge and chemical potential on brane could be given from the corresponding thermodynamic quantities of AdS dilaton black hole. With these thermodynamic quantities on brane, an extended Cardy-Verlinde formula is obtained in the large $r(\tau)$ limit. Meanwhile, our result has a similiar mathematical form with the one obtained in background of AdS-RN black hole \cite{Cai:2001jc}. This similarity is reasonable in physics, when making the dilaton coupling constant $\alpha=0$, the charged AdS dilaton black hole solution will return the AdS-RN one. Thus, when brane approaches the AdS boundary, we naturally obtain a similiar Cardy-Verlinde formula on brane like the one in \cite{Cai:2001jc} but with some corrections of dilaton field.

As discussions, in light of this work, we point out that the following extensions are still worthwhile to explore. The first one is, investigate the holographic complexity growth in the FLRW brane and black hole system constructed in this paper. Besides, as indicate in \cite{Kawai:2015lja}, in frame of AdS/CFT correspondence, the reheating process in brane-universe could be dual to the collapse of a spherical shell and formation of a black hole in AdS bulk. Thus, a valuable work is to study the holographic reheating in current brane-bulk system. What's more, it is an interesting topic to consider the movement of brane in background of asymtotically (A)dS black hole with scalar and electromagnetic hair  \cite{Brihaye:2019gla,Ferreira:2017cta,Herdeiro:2016plq}, and check that how the Cardy-Verlinde formula is affected by the scalar charge. Finally, it is also worthwhile to explore the movement of brane in the inner spacetime of some non-singularity black hole solutions \cite{Zeng:2018pzk,Zeng:2018kbv,Zeng:2016epp,Glavan:2019inb}.
\\
\section{Acknowledgements}
AC is grateful for Lei-Hua Liu and Carlos Herdeiro of the fruitful discussions and comments for this research topic and manuscripts, and thanks Ru-Yong Li for her help in adjusting the physical parameters in a picture. This work is supported by NSFC grant no.11875082. Besides, this work is also supported by The Center for Research and Development in Mathematics and Applications (CIDMA) through the Portuguese 
Foundation for Science and Technology (FCT - Fundação para a Ciência e a Tecnologia), references UIDB/04106/2020 and UIDP/04106/2020.

\end{document}